%% file: main-2025-3-7-exact.tex
\documentclass[11pt,a4paper]{article}
\usepackage{jheppub}

\input{preamble}

\title{Analytical Solution of the  Nonlinear Relativistic Boltzmann Equation}

\author[a]{Jin Hu}

\affiliation[a]{Department of Physics, Fuzhou University, Fujian 350116, China}

\emailAdd{hu-j23@fzu.edu.cn}

\cout{

\author{Jin Hu}
\affiliation{Department of Physics, Fuzhou University, Fujian 350116, China}

}

\abstract{
 We provide an exact analytical solution to the nonlinear relativistic Boltzmann equation for  a homogeneous, anisotropically scattering massless gas.  Utilizing  a BKW-like  trial solution, we cast the Boltzmann equation into a set of nonlinear coupled equations for scalar moments, based on which the analytical solution is derived.  One remarkable feature of our analytical solution lies in the nontrivial scattering angle dependence. 
  We also show that  this analytical solution  admits a stable fixed point corresponding to the equilibrium solution  as long as the parameters are physically feasible.  Furthermore,  a clear correspondence between our solution and the BKW solution pertaining to nonrelativistic Maxwell molecules is established, thereby clarifying the non-existence of a BKW-type solution in the relativistic domain for  massive particles. 

} 

\begin{document}

\maketitle

\section{Introduction}\label{s1}

The relativistic Boltzmann equation serves as an essential instrument for detailing the dynamics of dilute gases as they evolve out of equilibrium, holding a significant place within the realm of relativistic kinetic theory. This equation holds  a profound significance from a physical perspective,
 and has been extensively applied across  various sectors of physics,  including its role in describing the quark-gluon plasma produced in high-energy heavy-ion collisions  \cite{Heinz:1984yq,Heinz:1985qe,Bass:1998ca,Arnold:2000dr,Molnar:2001ux,Xu:2004mz,Denicol:2012cn}, and its utility in the theoretical exploration of cosmological phenomena \cite{Dodelson:2003ft,Weinberg:2008zzc}. Despite the extensive research performed on the Boltzmann equation across an array of physical contexts, finding analytical solutions to its intricate integro-differential structure remains a challenge, with such solutions being few and far between—even in nonrelativistic scenarios. A particularly distinguished analytical solution is that of Bobylev \cite{Bobylev1}, and the concurrent contributions  of Krook and Wu (referred to as BKW) \cite{KW,KW2}. This solution, originally discovered by Krupp in his dissertation \cite{dissertation} and now recognized as the BKW solution, as detailed in the review \cite{exactreview}, delves into the nonlinear relaxation dynamics of a nonrelativistic homogeneous gas. The gas is characterized by an elastic cross-section that varies inversely with the relative velocity of the colliding particles. The BKW solution offers a wealth of knowledge regarding the nuances of nonuniform relaxation processes, which are governed by the nonlinear collision term embedded within the Boltzmann equation. It is especially enlightening in its elucidation of the behavior of the one-particle distribution function at the high-momentum tail of the spectrum. 

Motivated by the pressing need to decipher the mechanisms of  thermalization and hydrodynamization in relativistic gas systems, and to assess the reliability and accuracy of numerical methods used in Boltzmann and hydrodynamic simulations, there has been an intense pursuit of exact analytical solutions. Yet, the complexity engendered by nonlinear collision integrals poses a formidable barrier to achieving analytical solutions for the relativistic Boltzmann equation, even under the simplest interaction models. A strategic bypass to this obstacle involves the use of a linearized Boltzmann equation; however, the linearized collision kernel is no less challenging to manage effectively.
Advancements have been made with the recent analytical determination of the eigenspectrum for the linearized Boltzmann collision operator within massless scalar $\phi^4$ theory \cite{Denicol:2022bsq}. Despite this progress, extending these findings to encompass more realistic interactions and the inherent nonlinearities remains a challenging task \footnote{Pioneering work on the analytical eigenspectrum was given by C.S. Wang and U.E. Uhlenbeck, focusing on Maxwell molecules in the non-relativistic regime (for details, refer to Chapter IV of \cite{deboer}).}. To tackle these difficulties and make progress, the Anderson-Witting equation, commonly known as the relaxation time approximation (RTA) \cite{1974Phy....74..466A} and firstly developed in non-relativistic kinetic theory called BGK model \cite{bgk}, has been broadly utilized. For an overview of ongoing efforts in this field, one may refer to \cite{Florkowski:2013lya,Denicol:2014xca,Denicol:2014tha,Nopoush:2014qba,Noronha:2015jia,Hatta:2015kia}, along with the cited references therein. It is essential to acknowledge that the utilization of the relaxation time approximation in the Boltzmann equation is  only theoretically justified under certain specific conditions related to the details of the interactions \cite{Hu:2024tnn}. Therefore, a careful examination of the eigenspectrum properties of the linearized collision operator is necessary to ensure the appropriateness of these results. This analysis is crucial for maintaining the integrity of the conclusions drawn from the use of such approximations.

The first analytical solution of the nonlinear relativistic Boltzmann equation is given for a massless gas with hard sphere potential in a homogeneous, isotropically expanding system \cite{Bazow:2015dha,Bazow:2016oky}. Advancing beyond the current state of research, we introduce a more sophisticated model that accounts for the differential cross section with a nontrivial  dependence on the scattering angle, expressed as $\sigma = \kappa \chi(\Theta)$. Here, $\Theta$ represents the  scattering angle as measured in the center-of-momentum reference frame, while $\kappa$ denotes the total cross section, which remains constant irrespective of the particles' momentum.  This parameterization of the cross section is justified when scattering processes dominate at a characteristic energy scale $T_{eff}$, allowing for a simplification where the momentum of  particles  can be effectively approximated by $T_{eff}$. 


The  paper is organized as follows.  In Sec.~\ref{part1}, we briefly review  the basic aspects of the  nonlinear Boltzmann equation, and introduce the moment method, from which one can construct a set of coupled equations for moments that is equivalent to the original Boltzmann equation. In Sec.~\ref{relax0},  we show how to solve this set of moment equations to obtain an exact analytical solution. In the following Sec.~\ref{relax1}, we discuss the parameter range where the obtained analytical solution respects physical consistency. We find that this analytical solution  admits a stable fixed point corresponding to the equilibrium solution  as long as the parameters are physically feasible.  Sec.~\ref{novel0} serves as a discussion on relativistic massive and non-relativistic transport. 
Summary and outlook are given in Sec.~\ref{su}.  

Natural units $k_B=c=\hbar=1$ are employed. The metric tensor  is given by $g^{\mu\nu}=\diag(1,-1,-1,-1)$ , while $\Delta^{\mu\nu} \equiv g^{\mu\nu}-u^\mu u^\nu$ is the projection tensor orthogonal to the four-vector fluid velocity $u^\mu$. The abbreviation dP stands for $\int dP\equiv \frac{2}{(2\pi)^3}\int d^4p\, \theta(p^0)\delta(p^2 - m^2)$.

 \section{Boltzmann equation and Moment method}
 \label{part1}
The space-time evolution of the non-equilibrium gases is described by the relativistic Boltzmann equation
\begin{align}
\label{Boltzmann}
&p\cdot\partial f(x^\mu,p^\mu)=C[f],
\end{align}
with
\begin{align}
 \label{Ckl}
 &C[f]
 \equiv\;\frac{1}{2}
 \int  {\rm dP^\prime}  {\rm dP_1} {\rm dP_2} \big(f(x^\mu,p^\mu_1)f(x^\mu,p^\mu_2)-f(x^\mu,p^\mu)f(x^\mu,p^{\prime\mu})\big) 
 W_{p,p^\prime\to p_1,p_2},
\end{align}
where $f(x^\mu,p^\mu)$ denotes the one-particle distribution function in  phase space, and $C[f]$ represents the  collision kernel. Our work is concentrated on local interactions involving two particles, excluding the external forces and quantum statistical effects. The transition rate is expressed as $W_{p,p^\prime\to p_1,p_2}=(2\pi)^6s\sigma(s,\Theta)\delta^{4}(p+p^\prime- p_1-p_2)$ with the differential  cross section $\sigma(s,\Theta)$ depending on  the total center-of-momentum energy squared $s\equiv (p^\mu+p^{\prime \mu})(p_\mu+p^\prime_\mu)=(p_1^\mu+p_2^{\mu})(p_{1,\mu}+p_{2,\mu})$, and $\Theta$, the scattering angle within the  center-of-momentum frame defined as $\cos\Theta\equiv \frac{(p_1-p_2)\cdot (p-p^\prime)}{(p-p^\prime)^2}$. This formula, Eq.\eqref{Boltzmann},  inherently accounts for the principle of detailed balance, $W_{p,p^\prime\to p_1,p_2}=W_{p_1,p_2\to p,p^\prime}$.
Note that Eq.\eqref{Boltzmann} is specific to one-component systems; for generalizations to multi-component systems, refer to \cite{DeGroot:1980dk, Hu:2022vph}.

As discussed earlier, the quest for an analytical solution to the Boltzmann equation is possible only  in limited special cases.  In the present study, we concentrate our focus on scenarios involving elastic scattering events characterized by a cross-section that does not vary with momentum, expressed as $\sigma\equiv \kappa \chi(\Theta)$.  Additionally, we define the total cross-section through the integral
    \begin{align}
    \sigma_T&=\frac{1}{2}\int d\Phi  d\Theta\sin\Theta\sigma(\cos\Theta),
    \end{align}
   where $\kappa$ is determined by ensuring that
    \begin{align}
    \label{nor}
    \frac{1}{2}\int d\Phi  d\Theta\sin\Theta\chi(\cos\Theta)=1.
    \end{align}   
   This enables us to regard $\kappa$  as   the total cross section $\kappa=\sigma_T$. It should be noted that the dependence on $\Theta$ is expressed in terms of  $\cos\Theta$ \cite{DeGroot:1980dk}. Generally, $\kappa$ may be contingent upon a characteristic energy scale that is relevant to the particular physical scenario under consideration, such as the temperature $T$ or the Debye screen mass $m_D$.
   
  To proceed, we impose the assumptions of spatial homogeneity and isotropy, which means that the one-particle distribution function is a function of time and the magnitude of momentum only, denoted as  $f=f(t,p^0)$.  We further  introduce scaled time $\tau\equiv T_0^3\sigma_T t$ and scaled momentum $\hat{p}^\mu\equiv p^\mu/T_0$ with an energy scale $T_0$ to transform  Eq.\eqref{Boltzmann} into a dimensionless equation,  
   \begin{align}
    \label{Boltzmann21}
   p^0\partial_\tau f(\tau,p^0)=C[f],
   \end{align}
   where the collision term is adjusted accordingly,
    \begin{align}
    C[f]&=\;\frac{(2\pi)^6}{2}\int  {\rm dP^\prime}  {\rm dP_1} {\rm dP_2}
       s\delta^{4}(p+p^\prime-p_1-p_2)\nn\\
    & \times \chi(\cos\Theta)  \big(f(\tau,p_1^0)f(\tau,p_2^0)-f(\tau,p^0)f(\tau,p^{\prime 0})\big).
    \end{align}
    Note that in the above two equations, $p$  now represents the scaled, dimensionless momentum, which applies similarly to other variables constructed by $p$). Hereafter, we will drop the hat notation from $\hat{p}$ for simplicity as long as nothing confusing occurs.      
       
  We employ the method of moments to address  Eq.(\ref{Boltzmann21}), which is transformed into a set of coupled equations for scalar moments
  \begin{align}
  \label{scalar}
  \partial_\tau \rho_n (\tau)=C^{(n)}(\tau),
  \end{align}
  with 
       \begin{align}
       &C^{(n)}(\tau)
       =\;
       \frac{(2\pi)^6}{2}\int  {\rm dP}{\rm dP^\prime}  {\rm dP_1} {\rm dP_2}
       \delta^{4}(p+p^\prime-p_1-p_2)\nn\\
       & \times s (p^0)^{n}\chi(\cos\Theta)  \big(f(\tau,p_1^0)f(\tau,p_2^0)-f(\tau,p^0)f(\tau,p^{\prime 0})\big).
       \end{align}
 and the definition for the energy moments
 \begin{align}
 \rho_n\equiv \int dP (p^0)^{n+1}f(\tau,p^0),
 \end{align}
 where $n$ is an integer value.
        
 Given that these energy moments act as  the relevant physical quantities, Eq.(\ref{scalar}),   as the evolution function of  scalar energy moments, can be regarded as encapsulating the identical physical essence as the original Boltzmann equation. Illustratively, the first two moments exhibit conservation properties, as evidenced by the equations
  \begin{align}
  \label{rho01}
  \partial_\tau \rho_0 (\tau)=\partial_\tau \rho_1 (\tau)=0,
  \end{align}
  where $\rho_0$ and $\rho_1$ correspond to the dimensionless energy and particle densities, respectively, with  $n\equiv \rho_0(\tau)$ and $e\equiv \rho_1(\tau)$. These densities, $n_0$ and $e_0$, are recognized as constants. The vanishing of the right-hand-side (RHS) of Eq.(\ref{scalar}) is a direct manifestation  of the collision invariance inherent to the Boltzmann equation.
  
  In our quest for an analytical solution, we propose a trial solution of the BKW form 
    \begin{align}
    \label{bkw}
f(p^0,\tau)=\lambda e^{-\frac{p^0}{\alpha(\tau)}}(A(\tau)+B(\tau)p^0), \end{align}
 with $\lambda$ symbolizing the fugacity, a factor that can be integrated into the definitions of $A(\tau)$ and $B(\tau)$. 
 The task then becomes the determination of $\alpha(\tau)$, $A(\tau)$ and $B(\tau)$ that are consistent with the moment equation Eq.(\ref{scalar}). 
  
  \section{ Exact analytical solution}
  \label{relax0}
 Given the trial solution $f(p^0,\tau)= e^{-\frac{p^0}{\alpha(\tau)}}(A(\tau)+B(\tau)p^0)$, we aim to seek the exact analytical solution to Eq.(\ref{scalar}). 
Substituting  the trial solution into the left-hand-side (LHS) of Eq.(\ref{scalar}) leads to
    \begin{align}
    \label{lhs}
    \text{LHS}&= -\frac{1}{2 \pi ^2} \Gamma (n+3) \alpha (\tau )^{n+2} \Big(\alpha (\tau) \big(-A'(\tau )-(n+3) \alpha (\tau ) B'(\tau)\,\big)\nn\\
    &-(n+3) \alpha '(\tau ) \big(A(\tau )+(n+4) \alpha (\tau ) B(\tau )\,\big)\Big), \end{align}
where the prime  denotes the derivative with respect to $\tau$. Turning our attention to the right-hand side, we derive the following expression
    \begin{align}
    \label{rhs}
    \text{RHS}&=-\frac{B^2(\tau)}{2}(2\pi)^6\int  {\rm dP}{\rm dP^\prime}  {\rm dP_1} {\rm dP_2}
    (u\cdot p)^{n} s\chi(\cos\Theta)\nn\\
    & e^{-\frac{u\cdot (p+p^{\prime })}{\alpha(\tau)}}\delta^{4}(p+p^\prime-p_1-p_2)(u\cdot p u\cdot p^{\prime }-u\cdot p_1 u\cdot p_2),
    \end{align}
    where $p^0$ is rewritten as $p^0=u\cdot p$ to  formally  restore the Lorentz covariance, which simplifies the application of Lorentz transformations in the computation of the collision integral. Within this context, $u$ represents a unit time-like vector, which  will be set to $(1,0,0,0)$ after finishing the calculation.

  For the calculation of the collision integrals, it proves to be convenient to transform the above expression in terms of the total and relative four-momentum 
    \begin{align}
    P^\mu&\equiv p^\mu +p^{\prime\mu}, \quad Q^\mu\equiv \Delta^{\mu\nu}_P(p_\nu-p^\prime_\nu),\nn\\
     Q^{\prime\mu}&\equiv \Delta^{\mu\nu}_P(p_{1,\nu}-p_{2,\nu}),\quad \text{with}\quad  \Delta^{\mu\nu}_P\equiv g^{\mu\nu}-\frac{P^{\mu}P^{\nu}}{P^2}.
    \end{align}
    The inverse transformation can be readily expressed as 
    \begin{align}
    p^\mu&=\frac{1}{2}(P^\mu+Q^\mu),\quad p^{\prime \mu}=\frac{1}{2}(P^\mu-Q^\mu), \nn\\
    p_1^\mu&=\frac{1}{2}(P^\mu+Q^{\prime\mu}),\quad p_2^{\prime \mu}=\frac{1}{2}(P^\mu-Q^{\prime\mu}).    
    \end{align}
         
Then  Eq.(\ref{rhs}) can be cast into the following form,
    \begin{align}
    \label{rhs1}
\text{RHS}
&=-\frac{B^2(\tau)}{2^{n+3}}(2\pi)^6\int  {\rm dP}{\rm dP^\prime}  {\rm dP_1} {\rm dP_2}
  \nn\\
 &\times \delta^{4}(p+p^\prime-p_1-p_2)s\chi(\cos\Theta)\nn\\
& \times e^{-\frac{u\cdot P}{\alpha(\tau)}}\left(\,(u\cdot Q^\prime)^2-(u\cdot Q)^2\right) (u\cdot P+u\cdot Q)^n\nn\\
&=\frac{B^2(\tau)}{2^{n+2}}\sum_{l=0}^{n}C^l_nI_{nl},
    \end{align}
where $C^l_n$ is the Binomial coefficient and $I_{nl}$ is given by
    \begin{align}
    I_{nl}\equiv &\frac{1}{2}(2\pi)^6\int {\rm dP}{\rm dP^\prime}  {\rm dP_1} {\rm dP_2}
    \chi(\cos\Theta) \delta^{4}(p+p^\prime-p_1-p_2)\nn\\
    &\times se^{-\frac{u\cdot P}{\alpha(\tau)}}\left(\,-(u\cdot Q^\prime)^2+(u\cdot Q)^2\right) (u\cdot P)^l(u\cdot Q)^{n-l}.
    \end{align}

    One can easily find that $I_{nl}$ is exactly a combination of Eq.(13) on Page 375 of the textbook on relativistic  kinetic theory \cite{DeGroot:1980dk} up to some trivial multiplicative factors. Putting it more specifically, we have
    \begin{align}
    \label{inl}
    I_{nl}=\frac{\alpha^{8+n}(t)}{4\pi^4}( -J^{(0,l,n-l,2,0)}+J^{(0,l,n-l+2,0,0)})
    \end{align}
    where
     \begin{align}
     \label{jab}
     &J^{(a,b,d,e,f)}\equiv  \frac{\pi  (d+e+1)\text{!!}}{2}  \sum _{g=0}^{\min (d,e)} \sum _{h=0}^{\left\lfloor \frac{b}{2}\right\rfloor }\sigma(f,g) \nn\\
     &\times \frac{(1+(-1)^{d-g}\,) (1+(-1)^{e-g}\,)d! e!}{4(d-g)\text{!!} (d+g+1)\text{!!} (e-g)\text{!!} (e+g+1)\text{!!}}\nn\\
     &\times \frac{(-1)^h b! (2 h-1)\text{!!}  2^{\frac{1}{2} (4 a+2 b+d+e+4 f-2 h+6)}}{(2 h)! (b-2 h)!}\nn\\
     &\times \Gamma (a+f+2) \Gamma \left(a+b+\frac{d}{2}+\frac{e}{2}+f-h+3\right),
     \end{align}
where $C^l_n$ represents the Binomial coefficient and $\sigma(f,g)$ denotes the Legendre expansion of the cross section $ \chi(\cos\Theta)$ weighted by $(\cos\Theta)^f$
\begin{align}
\label{sfg}
\sigma(f,g)\equiv \frac{2g+1}{2}\int_{-1}^1dx\, x^f P_g(x) \chi(x).
\end{align}
Here $P_g(x)$ is the Legendre polynomial. While an exhaustive derivation of the aforementioned expression is not presented here, we recommend the readers to refer to Chapter XIII in \cite{DeGroot:1980dk},  or the appendices in \cite{Hu:2022vph} for more technical details.
    
    The collision invariance is easily confirmed for the cases when $n=0$ and $1$ by examining Eq.(\ref{inl}). Furthermore, it is essential that Eq.(\ref{rho01}) is also met, which results in the derivation of the following expressions
    \begin{align}
    \label{AB}
    A(\tau )=\frac{\pi ^2 (4 n_0 \alpha (\tau )-e_0)}{\alpha (\tau )^4},B(\tau )= \frac{\pi ^2 (e_0-3 n_0 \alpha (\tau ))}{3 \alpha (\tau )^5}.
    \end{align}
    Consequently, the functions $A(\tau)$ and $B(\tau)$ are expressible in terms of  $\alpha(\tau)$, thus narrowing our computational focus to determining $\alpha(\tau)$.
    
    By substituting  Eq.(\ref{AB}) into Eqs.(\ref{lhs}) and (\ref{rhs1}), we formulate an equation designed to solve for  $\alpha(\tau)$, which is anticipated to be universally applicable for any value of $n$.  Guided by this insight, we strategically address the scenario where $n=2$, subsequently verifying whether the  solution is independent of $n$. Adhering to this strategy, we arrive at the equation governing  $\alpha(\tau)$,
    \begin{align}
    \label{alpha}
     \alpha '(\tau )=-\frac{1}{90} (2 \pi  \sigma (0,2)-5) (e_0-3 n_0 \alpha (\tau )).
    \end{align}
    This equation and Eq.(\ref{AB})  constitute the main results of this work. After substituting Eqs.(\ref{alpha}) and Eq.(\ref{AB}) back into Eqs.(\ref{lhs}) and (\ref{rhs1}),  it becomes evident that the resulting expressions for both the left-hand side (LHS) and the right-hand side (RHS) are composed of two separate multiplicative components: one that relies solely on $n$, and another that encapsulates the complete dependence on $\alpha(\tau)$. Furthermore,  the $\alpha$-dependent  parts are found to be in agreement  across both  LHS and RHS, regardless of the value of $n$. It is further confirmed  that the $\alpha$-independent parts are also consistent with  each other \footnote{The $\alpha$-independent parts within the series summation  in Eq.(\ref{rhs1}) elude analytical determination without  specifying the value of $n$. We analytically verified  that the first $5000$ terms match the LHS in Eq.(\ref{lhs}). }. 

\begin{figure}
    \centering
    \includegraphics[scale=0.5]{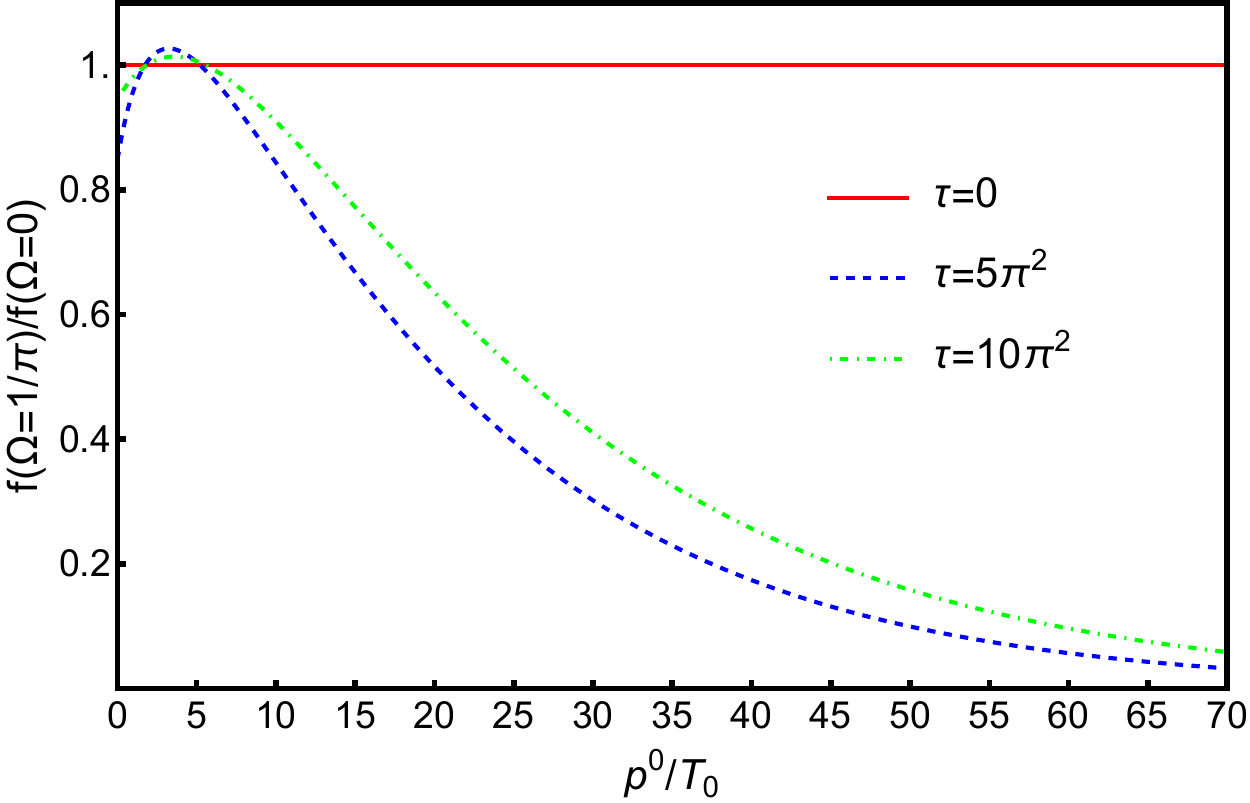}
    \caption{The ratio of the BKW solution for $\Omega\equiv\sigma(0,2)=1/\pi$ to the solution in the isotropic case.}
    \label{fig1}
\end{figure}
    
\begin{figure}
    \centering
    \includegraphics[scale=0.5]{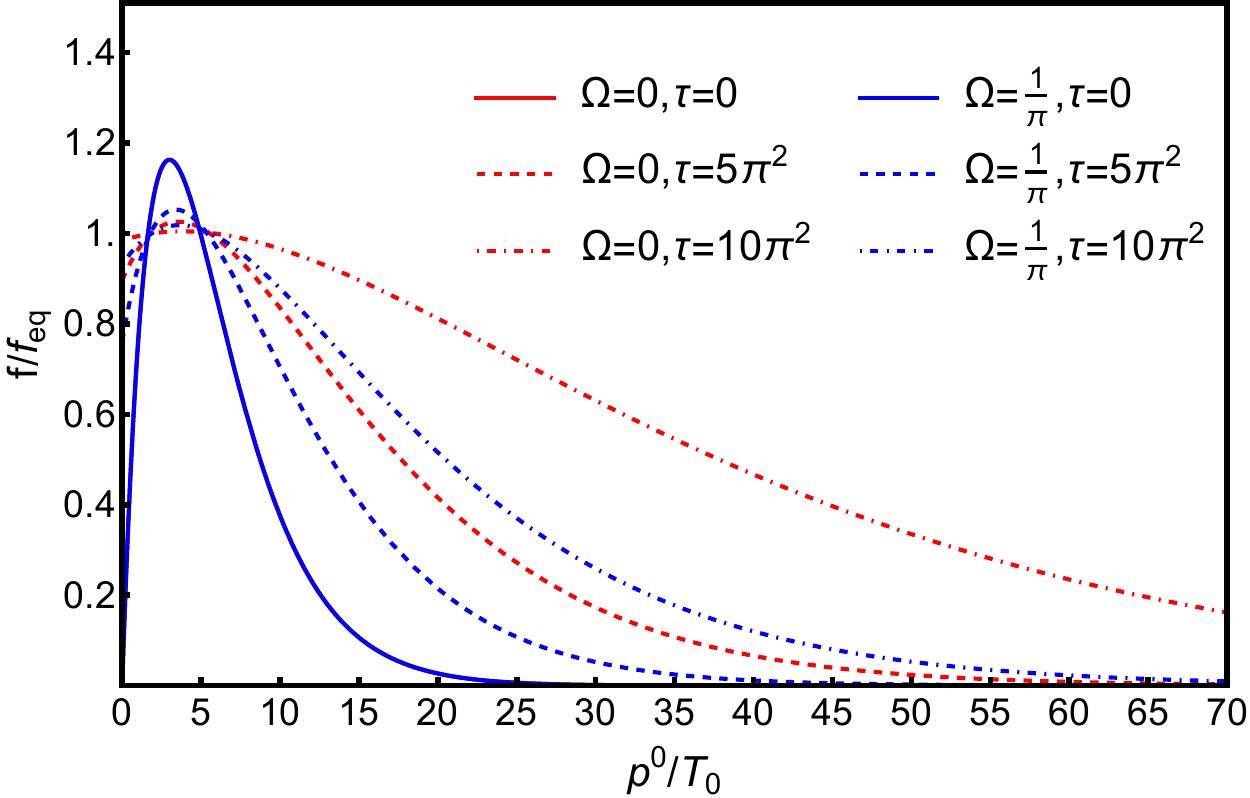}
    \caption{ The ratio of the BKW solution for $\Omega\equiv\sigma(0,2)=1/\pi$ and the isotropic case to the equilibrium distribution.}
    \label{fig2}
\end{figure}

    Without losing generality, we solve Eq.(\ref{alpha}) under the initial condition $\alpha(0)=\gamma$ to get
    \begin{align}
   \alpha(\tau)=( \gamma-\frac{e_0 }{3 n_0})e^{\frac{1}{90} \tau  (6 \pi  n_0 \sigma (0,2)-15 n_0)}+\frac{e_0 }{3 n_0}.
    \end{align}
    By setting $\gamma=\frac{3}{4}$, our solution for $f$  reduces to that given  in \cite{Bazow:2015dha,Bazow:2016oky}  in the Minkowski metric, where $\alpha(\tau)$ is replaced by $K(\tau)\equiv 1-\frac{1}{4}e^{-\tau n_0}$ defined therein. When the scattering cross section takes the isotropic
    limit, $\sigma(0,2)$ would vanish accordingly resulting in $ \alpha(\tau)=( \gamma-\frac{e_0 }{3 n_0})e^{-\frac{1}{6} \tau  n_0}+\frac{e_0 }{3 n_0}$.  To reflect the influence of anisotropic scattering on the system's evolution behavior, we present in Fig.\ref{fig1} the ratio of the BKW solution for $\sigma(0,2)=1/\pi$ to the solution in the isotropic case, where $\sigma(0,2)=1/\pi$ corresponds to $\chi(\cos\Theta)=\frac{3\cos^2\Theta}{2\pi}$ consistent with Eq.(\ref{nor}). It can be observed that, when time is sufficiently large, the two solutions exhibit significant differences. Additionally, Fig.\ref{fig2}  shows the ratio of the BKW solution for $\sigma(0,2)=1/\pi$ and the isotropic case to the equilibrium distribution. The results for the isotropic case are almost consistent with those presented in Ref.\cite{Bazow:2015dha} both in terms of the shape of the curves and the magnitude of the values, with minor differences attributed to the fact that the results in \cite{Bazow:2015dha} are obtained in the context of an FLRW expanding spacetime. In making the comparison, one should be aware that our definition of $\tau$ is different from \cite{Bazow:2015dha}.

  \section{ Physical solution and  fixed point}
  \label{relax1}
  
  We anticipate that within the realm of a physical system, the one-particle distribution function $f(\tau)$ remains non-negative  across all times where $\tau\geq 0$. Furthermore, it is essential that $f$ remains bounded as  $p^0$ tends towards infinity, which in turn places dynamic limitations on the form of the cross-section $\chi(\cos\theta)$. To articulate these physical prerequisites explicitly, we establish the following conditions
\begin{align}
\label{cross}
\sigma (0,2)  < \frac{5}{2\pi},\\
\label{ga}
 \frac{e_0}{4n_0}\leq \gamma \leq \frac{e_0}{3n_0}.
\end{align}
The first condition constrains the permissible forms of  the cross section $\chi(\cos\theta)$. Noncompliance with this stipulation would render the model incapable of harboring a physically viable analytical solution.  Moreover, the second condition is pivotal for defining the admissible range of initial conditions. For example, the solution in \cite{KW,KW2}  is deemed physically infeasible due to the possibility of  $f$ becoming negative \footnote{To address this issue, the authors of \cite{KW,KW2}  impose limitations on $\tau$, effectively adjusting the initial temporal setting to modify the initial condition. For  clarity, we advocate for the employment of  Eq.(\ref{ga}) to regulate the solution.}. Such physically inconsistent outcomes arise from the inappropriate selection of an initial condition, specifically  $\gamma=0$ (one should transform to the basic settings in \cite{KW} and then make judgments). Besides, the researchers in \cite{Bazow:2015dha,Bazow:2016oky} have opted for the minimum threshold of the physically permissible domain as defined by Eq.(\ref{ga}), thereby establishing  $\gamma=\frac{e_0}{4n_0}$. It is interesting to note $\alpha(\tau)$, if interpreted as the effective temperature out of equilibrium, is also bounded and saturated by $\frac{e_0}{3n_0}$
due to Eqs.(\ref{cross}) and (\ref{ga}).

As delineated by Eq.(\ref{alpha}), the solution admits a fixed point $\alpha=\frac{e_0}{3n_0}$. Assuming that Eq.(\ref{cross}) is satisfied, this fixed point is confirmed to be stable. Indeed, the asymptotic behavior of $\alpha(\tau)$ aligns precisely with this  fixed point. From this perspective, the analytical solution we have derived is of a transient nature, with an ultimate inclination towards this unique fixed point.  It is noteworthy that by equating $\frac{e_0}{3n_0}$ to unity, in accordance with  the matching condition  prescribed in \cite{Bazow:2015dha}, the fixed point corresponds to the equilibrium solution of the Boltzmann equation.

There is an additional remark on this analytical solution. Even though our treatment involves cross sections that describe anisotropic scattering processes,  the solution preserves the distribution isotropy in momentum space. At first sight, it might be confusing. However, it's important to recognize that ultraviolet (UV) physics is independent of infrared (IR) physics. The distribution function, as a statistical object, should be seen as part of IR physics, whereas the microscopic scattering cross section is inherently associated with  UV dynamics. Consequently, the anisotropy in the distribution function is unrelated to  the anisotropic scatterings. Typically, the anisotropic distribution originates from the macroscopic settings. For example, the initial distribution function lacks isotropy or an external field breaks the rotation symmetry. Once an isotropic initial condition is imposed, our analytical solution becomes applicable for characterizing the evolution that follows.
Generally,  the initial distribution of particles exhibits anisotropy in momentum space. In such circumstances, additional anisotropic irreducible moments must be also considered.
  
 \section{Relation with massive and non-relativistic transport}
 \label{novel0}
  
 In this section, we aim to elucidate the relation of our analytical solution with both massive and non-relativistic transport. To that end, we momentarily  focus on the Boltzmann equation for massive particles ($m$ denotes the mass) and write it in a form completely similar to the non-relativistic Boltzmann equation. Therefore, the  Boltzmann equation can be cast into the following three-notation form (see eq.(40) on Page 22 in \cite{DeGroot:1980dk})
\begin{align}
\label{Boltzmann1}
&\partial_t f=\frac{1}{2}
\int  d^3p^\prime d\Omega  \big(f(t,p^0_1)f(t,p^0_2)-f(t,p^0)f(t,p^{\prime 0})\big) 
v_{m}\sigma(s,\Theta),
\end{align}
where $d\Omega$ is a Lorentz scalar and can be regarded as an element of solid angle with respect to the center-of-momentum system, and $v_m$ is called  the Moller velocity
\begin{align}
&v_m\equiv \frac{F}{p^0p^{\prime 0}}=\bm{\sqrt{(v-v^\prime)^2-(v\times v^\prime)^2}},\quad \\
& \bm{v}\equiv \frac{\bm{p}}{p^0},\quad F\equiv \sqrt{(p^\mu p^\prime_\mu)^2-m^4}.
\end{align}
Note that $v_m$ is not a Lorentz scalar.

Let's first concentrate on its non-relativistic limit $p^\mu\rightarrow m(1,\bm{v})$. In this limit,
\begin{align}
v_m\rightarrow |\bm{v-v^\prime}|,\quad \cos\Theta\rightarrow \frac{(\bm{v-v^\prime)}\cdot \bm{(v_1-v_2)}}{(\bm{v-v^\prime})^2}.
\end{align}
 Thus, we obtain the non-relativistic Boltzmann equation
\begin{align}
\label{Boltzmann2}
\partial_t f&=\frac{1}{2}
\int  d^3\bm{v^\prime}d\Omega  \big(f(t,v_1)f(t,v_2)-f(t,v)f(t,v^{\prime })\big) \nn\\
&\times |\bm{v-v^\prime}|\sigma(|\bm{v-v^\prime}|,\cos\Theta),
\end{align}
where the conservation of energy-momentum is implied in the above equation.

In formulating Eq.(\ref{Boltzmann2}), we replace $p^0$ with $\frac{1}{2}mv^2$ in the distribution function $f(t,p^0)$, and  suppress explicit dependence on $m$. Concurrently,  the $m^3$ factor originating from $d^3p_1$ has been absorbed into the definition of $f(t,v)$.  Furthermore, $s$ can be constructed using $|\bm{v-v^\prime}|$ and $m$ in this limit, leading us to substitute  $s$ with $|\bm{v-v^\prime}|$ (again, we omit the $m$-dependence).
When seeking the BKW solution, our attention is drawn to the Maxwell molecules, i.e., $|\bm{v-v^\prime}|\sigma(|\bm{v-v^\prime}|,\cos\Theta)\sim \chi(\cos\Theta)$, which precisely corresponds to the model discussed by Bobylev \cite{Bobylev1},  Krook and Wu \cite{KW,KW2}
\begin{align}
\label{Boltzmann3}
\partial_t f&=\frac{1}{2}
\int  d^3\bm{v^\prime}d\Omega  \big(f(t,v_1)f(t,v_2)-f(t,v)f(t,v^{\prime })\big) \chi(\cos\Theta).
\end{align}

Then let's proceed  to the ultra-relativistic case, where $m=0$. In this scenario, the Moller velocity turns into $v_m=1-\cos\theta$ with $\theta$ being the angle between $\bm{v}$ and $\bm{v}_1$. Furthermore, $v_m$ can be integrated out in the angle integration of $d^3p^\prime$, yielding a trivial multiplicative factor. Assuming that the cross section is independent of $s$, i.e., $\sigma\equiv \kappa \chi(\cos\Theta)$, Eq.(\ref{Boltzmann1}) reduces to the equation analytically treated in this script. The striking similarity  between the BKW model presented in Eq.(\ref{Boltzmann3}) and our model  becomes readily apparent, offering a clear rationale for the feasibility of deriving an analytical BKW-like solution in the ultra-relativistic regime. It is also noteworthy to mention that the hard sphere approximation, while seemingly straightforward, is conceptually distinct when applied to non-relativistic versus relativistic scenarios. In the instance where the angular dependence characterized by  $\Theta$  in the cross-section is neglected, the ultra-relativistic hard-sphere gas system gives rise to a solution of the BKW type \cite{Bazow:2015dha,Bazow:2016oky}, which more closely resembles  the behavior of non-relativistic Maxwell molecules rather than hard-sphere molecules. 


Lastly, we turn our attention to the issue of massive transport  between the two scenarios mentioned above. Note that $\sigma(s,\Theta)$ is a Lorentz-invariant object constructed from the two Lorentz scalars $s$ and $\cos\Theta$, implying that  $\sigma(s,\Theta)$ and $v_m$ cannot be combined to form a momentum-independent quantity (up to a function of $\cos\theta$). Consequently, we conclude that an analog of the BKW solution is not feasible for a relativistic massive gas  (we also confirm this using the moment method).



 \section{Summary and outlook}
 \label{su}
 In this study, an exact analytical solution to the nonlinear Boltzmann equation is presented for a massless gas characterized by spatial homogeneity and momentum isotropy. To the best of our knowledge, this is the first analytical solution to the nonlinear relativistic Boltzmann equation with non-isotropic scatterings. Furthermore, we demonstrate that this analytical solution with physical parameters specified by Eqs.(\ref{cross}) and (\ref{ga})  admits a stable fixed point, corresponding to the equilibrium state. Note that our model is based on the approximation  that the scattering cross section is independent of particle momentum, nontrivially extending the related studies using hard sphere approximation \cite{Bazow:2015dha,Bazow:2016oky}. In scenarios where momentum can be estimated  by some typical physical energy scales, our solution  captures the dynamics of non-equilibrium evolution in dilute gases. It is worthwhile to note that the integration of the Friedmann-Lemaitre-Robertson-Walker (FLRW) geometry \cite{weinberg} into our theoretical study is quite straightforward. Also, we show the relation between our model and the nonrelativistic BKW model, from which one can conclude that a BKW-like solution is not allowed for a relativistic massive system.
 
 Future research building upon this study can be envisioned in several directions.  Firstly, the novel analytical solution we have developed could serve as a rigorous benchmark for assessing the numerical accuracy of algorithms designed to tackle the relativistic Boltzmann equation as well as hydrodynamic equations. For instance, extending our current research to a multi-component system could offer an approximate yet feasible approach to understanding the dynamic evolution of quark-gluon plasma (QGP) formed in heavy-ion collisions. To achieve this, it would be sensible to model the momentum dependence using relevant physical scales and to apply isotropic initial conditions. By adopting this approach, one could elucidate the distinct roles that various scattering processes with nontrivial angular dependencies play in affecting the system's evolution behavior. Secondly,  although there are numerous numerical frameworks for solving the relativistic Boltzmann equation, a formally simple yet physically motivated analytical solution is expected to provide deep insights into the understanding of non-equilibrium phenomena such as equilibration, the emergence of hydrodynamic behavior, and high-momentum nonequilibrium tail of the distribution function.  It is our aspiration that  our results will significantly contribute to the field. 
 Thirdly, there is potential to explore more meaningful analytical solutions that build upon the foundation we have established. This could involve relaxing certain assumptions  to enhance our understanding. For example, the study performed here can be extended  to systems that exhibit momentum anisotropy or spatial inhomogeneity. An immediately feasible and promising direction is to analyze the symmetries of the nonlinear relativistic Boltzmann equation. By identifying appropriate symmetry transformations, we can potentially transform existing BKW solutions into other nontrivial physical solutions. For instance, we could explore whether it's possible to map the solutions presented in this work to the solutions of the expanding system described in Ref.\cite{Bazow:2015dha}, relying solely on symmetry analysis. Historically, the discovery of the BKW solution sparked a surge of interest among scientists in seeking exact analytical solutions to the Boltzmann equation. We also hope that our work can serve as a catalyst, promoting further research into the analytical structure and solutions of the relativistic Boltzmann equation and kinetic theory. Last but not least, it would be interesting to perform a stability analysis on this analytical solution to ascertain whether it displays characteristics of an attractor, which would further validate its relevance and robustness. Typically, the (linear) stability analysis around the BKW solution would lead us to the eigenspectrum problem of the linearized collision operator (denoted as $L_1$). Note that if the background is the global equilibrium configuration, the resulting linearized collision operator $L$ is self-adjoint and non-negative \cite{DeGroot:1980dk}, which ensures the stability of equilibrium. However, the property of $L_1$ and its eigenspectrum necessitates a comprehensive and technically detailed analysis, which is left to  future work.

\section*{Acknowledgments}
 J.Hu is grateful to Shuzhe Shi, Qiuze Sun and Yi Yin for helpful discussions.



\bibliographystyle{JHEP}
\bibliography{ref}

\end{document}

%% file: preamble.tex
\usepackage{graphicx}
\usepackage{latexsym}
\usepackage{amsfonts}
\usepackage{amssymb}
\usepackage{amsmath}
\usepackage{bm}
\usepackage{mathrsfs}
\usepackage{slashed}
\usepackage{comment}

\usepackage{dcolumn}

\allowdisplaybreaks[3] 

\usepackage[utf8]{inputenc} 

\newcommand{\cout}[1]{ \if 0 {#1} \fi }
\newcommand{\nn}{\nonumber}

\newcommand{\diag}{ {\rm diag} }

\usepackage{color} 
\usepackage[normalem]{ulem}  
